# THE CHARACTERISTICS OF ENJOYABLE ONLINE LEARNING FOR CULINARY ARTS STUDENT

by.
Endang Mulyatiningsih [a)], Sri Palupi [b)], Prihastuti Ekawatiningsih [c)], Ambar Rizqi Firdausa [d)]

[a,b,c,d] *Culinary Art Program, Faculty of Engineering, Universitas Negeri Yogyakarta. Indonesia*

Abstract

The Covid-19 pandemic outbreak has forced all courses to be carried out online, but only a few truly fulfill student expectations. This study aims to explain the characteristics of enjoyable online learning based on the platform, the content, and the learning model. Data collected using closed and open questionnaires which were responded to by 110 students in 2019/2020. The closed questionnaire revealed the most preferred learning elements, and the open questionnaire was to clarify their reasons. The data were arranged sequentially from the quantitative data to the qualitative data. The results of the study showed that (1) the preferred online platforms were Moodle, Google Meet, and WhatsApp. They like Moodle because the content is well structured, Google Meet is easily accessible, and WhatsApp is their daily routine application; (2) The learning content consists of 2 to 3 resources i.e.: 6-10 pages papers, 11-15 pages PowerPoint and 6-10 minute videos. Too much content causes a heavy learning burden; (3) Most students preferred the blended learning strategy. The synchronous lectures for 60-75 minutes can motivate them because they can interact with lecturers and other students. Asynchronous lectures are more flexible that can be done anytime and anywhere so that the students become more independent in their learning.

**Keywords**: online learning, learning models, blended learning, platforms online, and content of program

## INTRODUCTION

The Covid-19 pandemic outbreak has changed the way of teaching and learning from offline to online. All lecturers and students are forced to use online learning technology in which each of them has different technological skills. Aini identified the barriers in online learning during the COVID-19 pandemic among students, teachers, and institutions, i.e., infrastructure and connectivity, e-learning systems, technology mastery, and self-management. Lecturers who already had a heavy workload should maintain students' active involvement, though they feel lonely when teaching using videos during online learning. Lack of financial support and the problem of behavioral change is another problem for the institution in this current condition (Aini, Budiarto, Putra, & Rahardja, 2020). The majority of students (63%) only have moderate commitment, and less committed at the most due to inadequate infrastructure and unstable internet signal (Abidi & Aldhalemi, 2020). After having online teaching and learning experiences during this pandemic, most students agree that online learning is still the best model in the new normal era (Ismaili, 2021) (Amir et al., 2020). The reasons to put forward online learning since they can learn anywhere or anytime, easily access material, and independently adjust the pace of learning (Michał Baczek, 2020), E-learning is considered efficient because it provides more time to study and review the materials (Amir et al., 2020). If the outcomes between online and offline learning are compared, the knowledge aspect is not significantly different but in the case of skills and social aspects, it is less effective (Michał Baczek, 2020).

E-learning can be done in a simple or sophisticated form. The simplest e-learning model is carried out by (1) delivering learning within an online platform; (2) participating in a virtual lesson at specific times; (3) completing assignments and submitting in digital repository; (4) monitoring student work through an online platform. A more sophisticated e-learning model can be in the form of (1) a digital platform using Google Meet/Classroom, Zoom, Skype, Google Hangouts, Moodle for materials access; (2) an assignment submission with various forms, such as picture messages and uploading it in email, Google Drive, OneNote, etc.; (3) an assignment review and feedback provision through the system; (4) a questions/problems post on discussion forums (e.g., Facebook Group, Google Group, Microsoft team). Lecturers check students' learning progress one by one, students' difficulties who cannot join the forum, provide feedback on their assignments, discuss difficult questions, and prepare the following assignment. *Adapted from resources shared by: https://instructionpartners.org/; Wisconsin DPI - https://dpi.wi.gov/ and the Indiana DOE - https://www.doe.in.gov/.*

The implementation of online learning right now relatively uses a simple model. The survey results from Fadhil A. Abidi found that (1) Most lecturers (88%) explained that the course regularly based on the schedule; (2) 66% of lectures are conducted on theoretical subjects and 34% are conducted for practical, video presentations, or laboratory experiments (Abidi & Aldhalemi, 2020). However, the level of student participation in online learning is still lacking. The students admitted that they can use all online learning services but only 10.5% take online classes, 7.8% participate effectively, and 4.5% submit assignments, and take exams (Mahyoob, 2020). The success and failure of online learning can be influenced by internal and external factors. Independent learning plays an important role in determining the success of online learning. Students who have high self-regulated learning show independent learning with a bigger chance of success (Albelbisi & Yusop, 2019). The external factors of technical support and instructional design have a significant influence on the use of LMS. If they find LMS is "easy to use" and "useful", then they will have positive intentions to use it (Alshammari, 2020). Clearly stated learning goals and expectations, provision of high-quality learning materials, and efforts to improve collaborative learning can contribute significantly to student achievement with the e-learning (Vo, Zhu, & Diep, 2020).

Qualified e-learning positively impacts students' satisfaction and learning effectiveness (W. Zhang & Cheng, 2012). The quality of e-learning services, academic engagement, and student satisfaction will influence student academic achievement (Satuti, Sunaryanto, & Nuris, 2020). The components for a minimum quality online program include general learning objectives, program content, learning activities, information resources, communication tools, use of media, learning processes in virtual classes, and assessment strategies (Marciniak, 2018). The quality of e-learning is assessed from planning, development, process, and product (planning, development, process, and product) (W. Zhang & Cheng, 2012). The planning is qualified, if it pays attention to the analysis of competency needs, curriculum, instructional objectives, and material structure. The development involves the instructional design, course material design, website design, and assessment plans. Meanwhile, the process should fulfill technical support, website program, learning interaction, flexibility, as well as evaluation of learning processes and outcomes. The literature review shows a reciprocal influence between the teacher's ability to develop learning content and student learning independence towards the success of e-learning. Both aspects of lecturers' ability for content development and students' autonomy will support each other to achieve successful e-learning. As mentioned by Arfan Shahzad that there is a positive relationship between service quality and information quality on user satisfaction that determines the success of the e-learning portal. (Shahzad et al., 2021). Instructor quality, course design, proper feedback, and student expectations provide a positive impact on student satisfaction and subsequently positively impact student performance (Gopal, Singh, & Aggarwal, 123AD). A good and complete service is positively correlated to the quality of the student experience (Malinovski, Trajkovik, & Vasileva-Stojanovska, 2018).

At the beginning of the Covid-19 outbreak, there were many challenges faced by lecturers and students. Lecturers are challenged to design, teach, and evaluate learning outcomes virtually. They should give a quick response by enhancing their digital literacy in operating, creating, uploading, and sharing good video content. On the other hand, students should have self-management, technological literacy, and high motivation to overcome their self-obstacles in using effective technology for learning and eliminate emotional discomfort due to loneliness (Rasheed, Kamsin, & Abdullah, 2020). In many cases, students still expect face-to-face learning to build an emotional connection between students and lecturers (Siswati, Astiena, & Savitri, 2020). Students are educational users who are expecting the best learning services, though it is online. Students can learn more quickly and deeply when they obtain positive support from teachers and other fellow students. According to Waterworth, enjoyable learning can be achieved by maximizing the personality, emotions, and senses of students to build more positive and lasting learning outcomes. The contradiction of enjoyable learning is anxiety, stress, confusion, intolerance, bullying, and inflexibility (Waterworth, 2020).

A long period of learning may cause mental and physical fatigue (burnout). A happy student, good physical condition, suitable learning environment, high learning motivation is the key component of good study. Students who are sad, demotivated, tired, and surrounded with noisy environment will require a big effort and drain students' energy to study (Kamsa, Elouahbi, & El Khoukhi, 2018). In the context of e-learning, a fun atmosphere is rather difficult to achieve because students feel isolated and 'lose' the interactions with their peers. Based on this current, this study tries to identify the learning elements that can enhance students' motivation in the Culinary Art program based on the platforms, the content, and the online learning models.

**RESEARCH METHOD**

The research employed a mixed-method by combining quantitative and qualitative research starting from design, data collection, and data analysis to explain enjoyable e-learning implementation. The sequential explanatory designs were used to explain the findings of quantitative data analysis, followed by interpretation of qualitative data (Creswell, 2012). The research was carried out from March 2020 to June 2020 at the Culinary Art Program, Faculty of Engineering, Universitas Negeri Yogyakarta. The population of this research was the students in the academic year of 2019 (semester 4) and 2020 (semester 2) as many as 110 people during this Covid-19 Pandemic outbreak. The samples were taken using a stratified random sampling technique with the following detail.

| Academic Year; Class | Population | Sampel |
|---|---|---|
| 2020 A | 40 | 30 |
| 2020 D | 41 | 30 |
| 2019 A | 37 | 25 |
| 2019 D | 39 | 25 |
| Total | 157 | 110 |

The data were collected from closed and open questionnaires using the Google form application. The closed questionnaire was to collect quantitative data to reveal the most preferred online learning platforms, content, and models. The open questionnaire was used to collect qualitative data to clarify the students' responses in the closed questionnaire. The guideline of the instrument is presented in the following Table 1.

Table 1. The guideline of the research instrument

| E-learning Component | Options |
|---|---|
| Platform | LMS (moodle, google classroom, edmodo, etc) |
|  | Synchronous lectures (zoom, google meet, skype) |
|  | Communication (WhatsApp, Microsoft team, telegram, Facebook, hangout) |
| Content | Module (the number of pages, the characteristics of the content of the module) |
|  | PowerPoint (the number of pages, the characteristics of the content) |
|  | Video (the duration, characteristics of content) |
|  | External sources |
| Model e-learning | Learning models (flipped, blended, heutagogy) |
|  | The contents of synchronous activities (material explanation, discussion, presentation of assignments/task, feedback) |

Data analysis was carried out gradually in several stages, namely: (1) calculating the amount and the ranking of the quantitative data to provide information on the type and volume of the most preferred e-learning content; (2) collecting and displaying qualitative data to know why the students choose e-learning components (platforms, content and learning models); (3) reducing qualitative data that have similar meaning; (4) rearranging the responses into coherent, systematic sentences that can represent all student answers.

**Results and Discussion**

**1. Online Platform**

There are 3 types of online learning platforms frequently used by the lecturers, namely: Learning Management System (LMS); synchronous model, and asynchronous communication model. The LMS platform that many students like the most was Moodle or Google Classroom (65.5%). The synchronous learning platforms was Google Meet/Zoom (70.9%) and WhatsApp groups as the social media platforms for communication (95.4%), respectively.

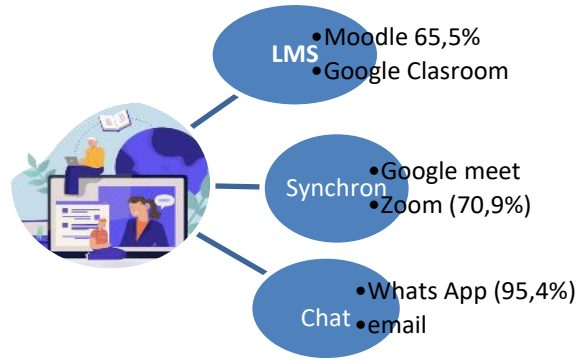

Figure 1. The preferred online learning platform

The reasons for choosing Moodle or Google Classroom are that (1) the information provided in Moodle is clearer, and the content, assignments, submission, and task deadlines in one semester are well structured; (2) the learning period in a week is more flexible that can be done anytime and anywhere; (3) learning content can be studied repeatedly that support independent learning, even the lecturers are still expected to explain the content through synchronous lectures; (4) Google Classroom is simpler than Moodle since the login does not need to use a university account, and the connection is faster and easier with a handphone. The use of quotas for asynchronous lectures using Moodle or Google Classroom is lower than synchronous lectures with Google Meet/Zoom.

The students prefer to the synchronous learning platform using Google Meet/Zoom for several reasons, such as: (1) the students and the lecturers can interact effectively where students can directly ask questions for any difficulties; (2) it can reduce the sense of isolation since they feel to meet face to face with their peers and hear the lecturers/friends' voices during the learning. Some students with the audio-lingual type find it easier to understand the material explained by the lecturer than studying on their own. Unfortunately, in synchronous lectures using Google Meet/Zoom, students seem to be not serious because they only face the laptop screen/ handphone and often turn off the camera. It is easy to access, but the signal is often constrained due to the unstable connection. Another problem is the high quota rate of asynchronous lectures. They also get confused if each lecturer gives many assignments.

WhatsApp groups are the only online platform that the students like the most to communicate, due to simple reasons. They always use WhatsApp in their daily activities, so they rarely miss the information from the group. This application can be used for chatting, calling, voice noting, sending files, taking photos, etc. but sending the tasks via email is prioritized due to the limited storage of WhatsApp. The consultation with WhatsApp chat is considered more polite than telephone because it does not require the lecturers to answer right away. The lecturers can reply at any time, and the chat can be known whether it has been sent or read. Moreover, this group application is easy to use for sharing information, materials, and group discussions.

**Content of Program**

Content of program is discussed from the number, type, and characteristics of learning resources based on the students' preference. The results of the quantitative analysis revealed that (1) the number of learning resources that were mostly chosen were 3 (36.7%) and 2 (32.11%) consisting of modules, power-points, and videos, respectively. The expected number of learning resources varied depending on students' learning abilities. The students with high learning abilities have no problem when they are given more than 3 learning resources because the material that does not exist in power-point can be found in modules, videos, and Google sources.

| Modul | PowerPoint | Video |
|---|---|---|
| • 6-15 page<br>• simple and completed with pictures | • 11-15 page<br>• not only textst | • 6-10 menit<br>• tutorial, completed with animation and text |

The more learning resources, the wider the knowledge insight will be gained if the learning resources are interesting and enjoyable. Students who are busy and lack time to study stated that the number of 2-3 types of learning resources is sufficient for learning because too many learning resources create confusion, and a learning burden, especially if the content is uninteresting. Some students even asked for only one source, but the complete one. They expect to access learning resources once and they can find all the needed materials and information. They are happy if each unit of learning material ends with an objective test to measure their learning achievement.

The preferred module characteristics include the number of pages in 6-10 pages (44%) and 11-15 pages (37.3%). A module of about 11-15 pages is sufficient to explain material with the following criteria i.e.: (1) complete material, adequate explanation data, and reference support; (2) clear instructions, summaries, and comprehensive quizzes; (3) simple and effective language which is easy to understand. The modules with more than 15 pages are less interesting for students because there are still many other sources to learn. The content of the module should be equipped with illustrations and examples to explain the difficult material (72.5%) since it (1) is easier to understand and interesting to read; (2) has a clear description; (3) has the complete and valid knowledge.

In the preferred PowerPoint (ppt) characteristics, 45.9% of students agreed that ppt pages could be more than modules around 11-15. The number of pages in ppt is not a big deal for students but there are some special requirements on the quality, such as (1) the interesting content, pictures, diagrams, voice notes, and photos to clarify the material in the module, not only text. (2) the explanation in the ppt should be concise, and straight to the core of the material or only important points so that it can be read faster than the module; (3) the content should be short, complete, and well-structured.

The characteristics of videos that students like the most are tutorials (77.1%) and animation (61.5%). It is better if the videos have a run time of approximately 6-10 minutes (59.6%). The qualitative data that explain these claims are: (1) the long video duration makes students bored and sleepy but too short video duration may not be completed so it is difficult to understand; (2) the long video duration is usually still watched but the time is accelerated and skipped in several parts; (3) the average concentration of student is in the first 10 minutes and decreases in the following minutes. Videos that are longer than 10 minutes should be cut into sections; (4) video duration that is too long is not efficient to be repeated; (5) short videos are expected to be better because the explanations are more concise, short, dense, clear, and easy to understand; (6) the internet quota to watch short videos is more efficient; (7) videos containing subject matter are often less interesting so that if the duration of the show is too long, it lower students' intention to access it. Moreover, interesting video content for Culinary Art students is a video tutorial with animations because it (1) contains systematic steps to create a product; (2) brings real practical learning situations that are easy to follow and replay.

**Learning Model**

Almost all students (93.6%) prefer the blended learning strategy as a balanced combination of synchronous and asynchronous lectures. The implementation of synchronous learning using the Google Meet/Zoom platform is 70.9% for 60-75 minutes (45%) and some even choose <60 minutes (39.4%). Synchronous learning focuses on the explanation of difficult materials (59.6%). Asynchronous learning uses the Moodle and Google Classroom platforms (65.5%) and communication utilize WhatsApp (94.5), respectively.

| Methods | Asynchronous activities | Asynchronous Media |
|---|---|---|
| • blended learning<br>• Intervensi individual | • video conference in 60-75 minutes<br>• the explanation of difficult materials | • tasks submission throuh LMS<br>• Communication with WhatsApp Group |

The Culinary Art program students give several reasons for selecting blended learning i.e. (1) the synchronous lectures can motivate because they can interact with lecturers and other students. (2) the asynchronous lectures are more flexible that can be done anytime and anywhere so that it supports independent learning; (3) some materials that can be studied by their own should be done asynchronously to be more efficient and effective. The qualitative data clarify the reasons why students choose a synchronous period between 60-75 minutes because too long synchronous (1) will make students bored, and decreases learning concentration so that the material presented by the lecturer cannot be well received by students; (2) is tiring because they have to sit continuously in front of the laptop even though the camera is usually off; (3) stimulates many tasks where they already had many assignments; (4) makes internet network unstable, the heat of handphone, runs out of battery backup and internet package, and other disturbances that are impossible to attend full-time lectures, especially if there is more than one course in a day, so many students leave meetings before the end of the lectures.

Synchronous lectures are still needed because 59.1% of students expect the lecturer to explain the material to avoid misconceptions. The material explanation from the lecturer is easier to understand than the peers' presentation. Lecturers should give brief and clear explanations completed with discussion sessions and humor to create a pleasant class atmosphere. The task feedback activity during synchronous lectures was highly appreciated by 32.7% of students. They want to know their capabilities and weaknesses that need to be improved. The qualitative data revealed that the lecturers of the Culinary Education Program still rarely provide feedback on their assignments. Some lecturers just give assignments without providing explanations which makes them confused because they do not know the right way to do it. Discussion and presentation activities in the synchronous model were mentioned by 23.6% of students. The reasons students choose this activity since (1) through presentation activities, students are encouraged to make interesting media, to learn the material deeply to be able to explain it to other students. This activity is also beneficial to share their experiences and it can train students' communication skills, critical thinking, and active involvement during the learning process.

**DISCUSSION**

The online learning platforms during the Covid-19 pandemic that are preferred by students consist of LMS (Moodle), video conference (Google Meet/Zoom), and chat/social media (WhatsApp). The preference of this platform is in line with previous research results. E-learning during the Covid-19 pandemic at Taibah University Saudi Arabia was carried out using WhatsApp to send and receive homework and assignments (72%), Email (53.60%), Zoom (33.50%) Google Classroom, and Microsoft team (24%) (Mahyoob, 2020);(Almusharraf & Khahro, 2020). Almusharraf adds that students are satisfied with Google Hangouts for face-to-face learning followed by Google Classroom and LMS (Moodle) for task management and assessment. Moodle (modular object-oriented dynamic learning environment) has several advantages over other LMS because it has more complete features and is safe to use. (Jebari, Boussedra, & Ettouhami, 2017).

The number of approved online learning content is 3 consisting of modules, presentation media (PowerPoint), and videos. Complete e-learning content may cause low student participation rates because the learning load is too heavy. The learning content that only contains simple modules, videos, and assignments can increase student learning participation by almost 100% (Mulyatiningsih, 2020). Students expect the number of modules and PowerPoint pages to be around 6-15 pages, but the content must be clear, interesting, and complete references. The poor quality of e-learning content may enhance students' knowledge but it does not contribute to behavioral change and skills development in the workplace (Alturkistani, Car, Wells, & Meinert, 2018). The preferred learning video should complete with interesting, creative, and unique animations as well as text narration. The duration of the selected video is around 6-10 minutes that support repetition. The results of this study are supported Afify's results which found that the students' performance who learn through short duration videos achieve better results on their cognitive achievement and retention of learning effects in the long term (Afify, 2020).

Blended learning is mostly chosen by the students since it provides a balanced composition between synchronous and asynchronous models. The students need synchronous learning because they can directly consult with the lecturers but if it is too frequent it may spend their internet quota. Meanwhile, asynchronous learning is considered more flexible and can be done anytime and anywhere, but asynchronous learning tasks are often too much so that it is very burdensome for students. Lengthy use of online interaction has revealed the many problems encountered by teachers and students. The content is abstract, many concepts exist that need real face to face interaction for complete understanding (Mishra, Gupta, & Shree, 2020). The flipped classroom and blended learning strategies have been proven effective for learning, and it has a positive effect on academic achievement and student engagement than the conventional methods (Talan & Gulsecen, 2019)**.** Students expect synchronous learning during the material explanation, task presentations, and assignment feedback. The material explained by the lecturer is considered easier to be understood than their peer presentation. The students hope their assignments are given feedback to eliminate mistakes on subsequent assignments. They think that individual guidance session can enhance their learning motivation, attitude, and self-efficacy than the classical session (J. H. Zhang et al., 2020).

## CONCLUSION

1. The students prefer a familiar and easy-to-use platform for their online learning platform. The results show that the students expect LMS (Learning Management System) using Moodle or Google Classroom (65.5%), synchronous learning platforms with Google Meet/Zoom (70.9%), and social media/communication utilizing WhatsApp groups (95, 4%). Moodle or Google Classroom provides clearer learning content, and it is well-structured. Google Meet/Zoom is considered easy to access but often arouses unstable signal problems. WhatsApp groups obtain the highest percentage because it is their daily application that they rarely miss information from it.
2. The online learning content expected by students is understandable, interesting, and brief. The modules, PowerPoint media, and videos are sufficient for learning with the following criteria (a) The number of module pages is about 6-10 pages (44%) and the content of the module is equipped with pictures and examples for the difficult material; (b) The number of PowerPoint (ppt) pages is more than the module so that the explanation is complete, adequate, and not only text; (c) The type of video tutorials should be equipped with creative animations with a running time of about 6-10 (59.6%) minutes.
3. Almost all students (93.6%) choose a blended learning strategy that combines synchronous and asynchronous processes. The synchronous lectures should be between 60-75 minutes to explain difficult materials, but the too long period will stimulate boredom, low concentration, and sleepiness so that the presented material will not be well-received by students. The students expect asynchronous lectures with small tasks portion since it is more flexible and can be done anytime and anywhere that support autonomous learning.

**Declaration of Competing Interests**
We declare that the publication of the results of this study is to help design joyful online learning.
**Recognition**
The researcher would like to thank Universitas Negeri Yogyakarta for facilitating the research group
**Funding**
Research funded by Universitas Negeri Yogyakarta, Indonesia

**Complete Data**

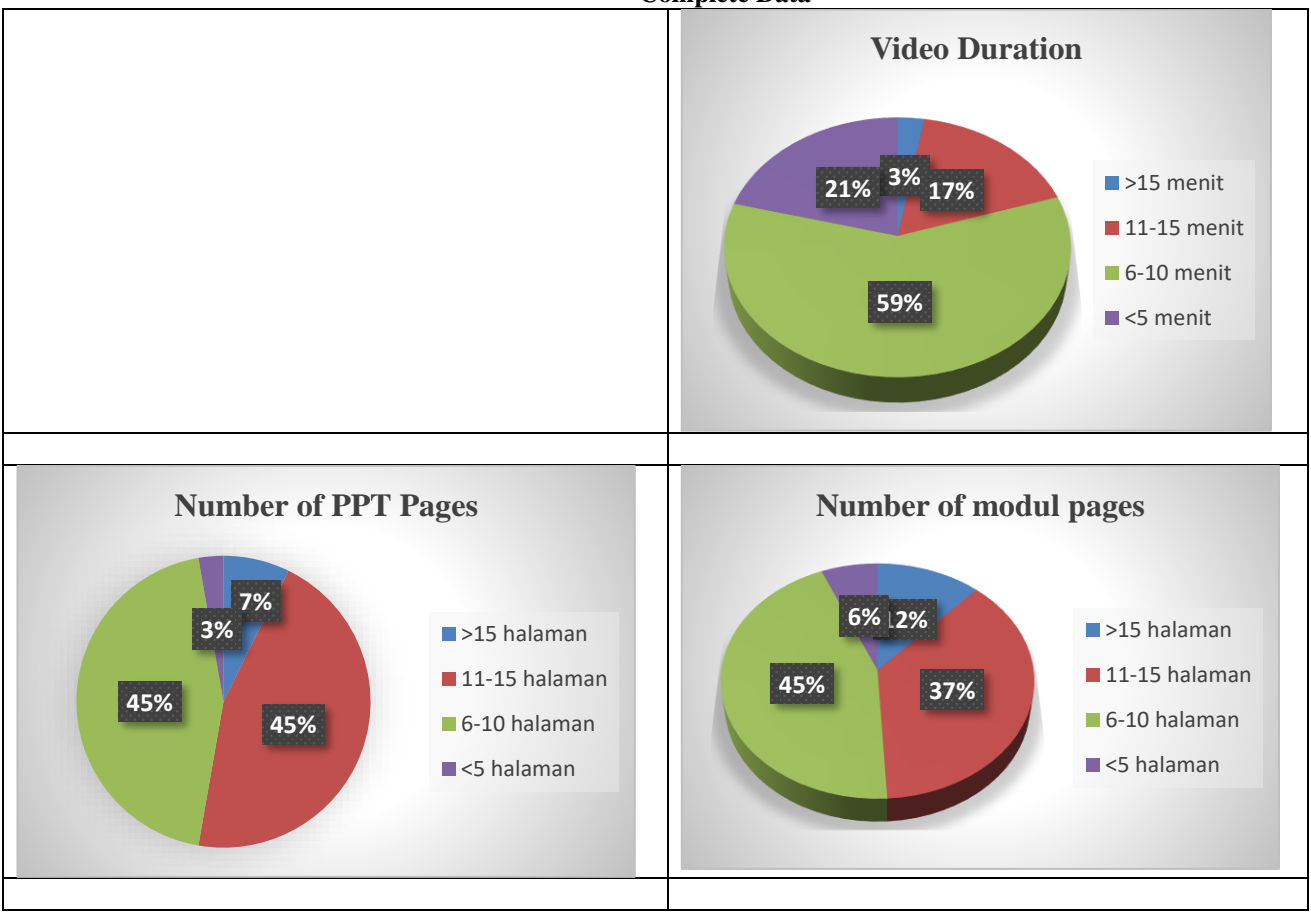

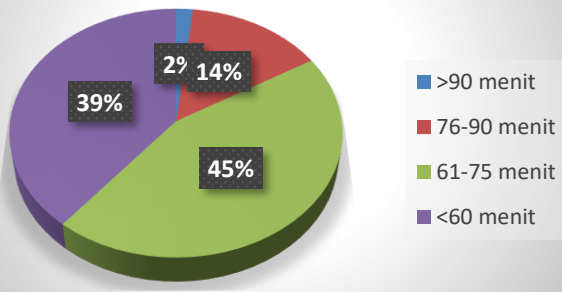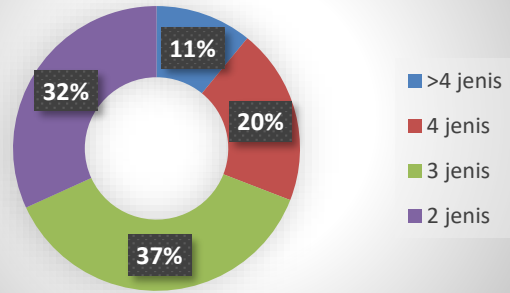